# POWER GATING STRUCTURE FOR REVERSIBLE PROGRAMMABLE LOGIC ARRAY

Pradeep Singla

Faculty of Engineering,
Asia-Pacific Institute of Information Technology, SD India

## ABSTRACT

*Throughout the world, the numbers of researchers or hardware designer struggle for the reducing of power dissipation in low power VLSI systems. This paper presented an idea of using the power gating structure for reducing the sub threshold leakage in the reversible system. This concept presented in the paper is entirely new and presented in the literature of reversible logics. By using the reversible logics for the digital systems, the energy can be saved up to the gate level implementation. But at the physical level designing of the reversible logics by the modern CMOS technology the heat or energy is dissipated due the sub-threshold leakage at the time of inactivity or standby mode. The Reversible Programming logic array (RPLA) is one of the important parts of the low power industrial applications and in this paper the physical design of the RPLA is presented by using the sleep transistor and the results is shown with the help of TINA- PRO software. The results for the proposed design is also compare with the CMOS design and shown that of 40.8% of energy saving. The Transient response is also produces in the paper for the switching activity and showing that the proposed design is much better that the modern CMOS design of the RPLA.*

## KEYWORDS

*Reversible programmable logic array, Sleep Transistors, Power gating, MUX gate, Feynman gate, Reversible logics, TINA- PRO*

## 1. INTRODUCTION

With the growing demands of the prominence portable systems, there are a rapids and innovative developments in the low power designs of the very large scale integration chips during the recent years [7]. The power dissipation or the energy consumption is one of the major obstacles for the Low power electronic design industries [4] [5]. For a low power digital hardware design, the researchers/ scientists are struggling for overcoming such a limiting factor/ obstacle and proposing different ideas and designing methods from the logical level ( Gate Level) to circuitry level (Physical level) and above[6]. Even after of such a struggle, there is no any universal method to design to avoid trade-off between delay, power consumption and complexity of the circuit. Still, the designer is required to opt. appropriate technology for satisfying product need and applications [6].

In case of the gate level implementation, reversible computing is one of the growing technologies for reducing the power dissipation [3]. The concept of the reversibility in the digital hardware design for the low power design basically came from the "Thermodynamics" which taught us the benefits of the reversible systems over the irreversible systems. The irreversible system design OR the system designed conventional approach consumes heat of equal to KTln2 on computation of every bit stated by Rolf Landauer, in 1961. Where K denotes the Boltzmann's constant which is equal to numerical value of $1.38 \times 10^{-23} m^2 kg^2 k^{-1} (J/K)$ and T is the temperature at which the logical computation is performed [1]. Even in 1973, Bennet correlate this heat loss with the information lost and stated the heat can be saved by using the reversible logics which does not dissipates heat [2]. So, this technique is best suitable for the logical / gate level implementation of the digital hardware. The ref. [3] showed the reversible design of the Programmable logic array called RPLA (Reversible programmable logic Array).

For a designing of the hardware, the circuit/ transistor/physical level implementation is further process of design. Today's Modern CMOS technology is used to design the system at transistor level and the scaling down of the feature size of the VLSI chips leads to improve the performance of the system. But, these technologies of the threshold voltage scaling down and decreasing feature size leads to the increment in the transistor leakage power with the exponential rate [9]. As the feature size becomes smaller, the length of the induced channel becomes shorter resultant sub threshold leakage current when the transistor is off.

Multi-Threshold CMOS OR Power gating technique is one of the prominent techniques by which the leakage current in the low power circuits in standby mode/ at the time of inactivity can be reduced[8]. In the design of power gating, multi-threshold CMOS has been introduce with the low & high threshold Voltage ( $V_{th}$) unit linked to the ground and to power supply respectively called sleeps transistors as shown in fig.1 [10]

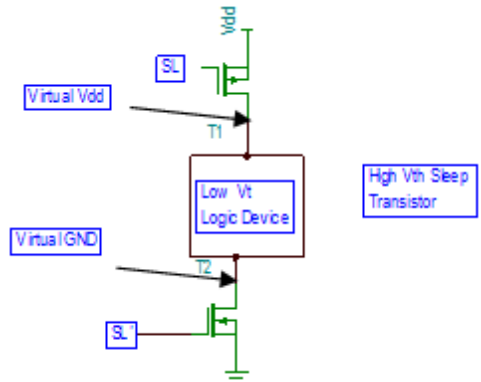

Figure1. Power gating structure

There are two operational modes of Multi-Threshold CMOS technique for the saving of power or reducing the power dissipation named active and sleep. In this technique, the low threshold voltage logic device from power supply and the ground via sleep transistor is used so, this technique is also known as power gating [10].

The Programmable logic array (PLA) is the one of the yardstick that provides customers a wide range of voltage characteristics, logic capacity & speed [10]. There of numbers of applications of PLA's like ultrasonic flow detection, DSP's etc. Even the PLA's are faster than the high speed DSP's [11]. The programmable logic array has many applications in the medical and industrial field also. In the Reference paper 3 written by the same author already showed the cost- effective design of RPLA for the low power design by using the reversible logics and ensures that design does not dissipates heat called Reversible programmable logic array (RPLA) [3]. So, with the huge numbers of benefits of the RPLA (Reversible Programmable logic array) design over the conventional PLA design, a power efficient design of reversible programmable logic array for the low power industrial applications by using the multi threshold distributed sleep transistor network at circuit level implementation is proposing in this paper. In order to make obvious the proposed architecture of power efficient RPLA, a 3- input RPLA which be capable of performing any $2^3$ functions using the combination of 8 min terms is also designed and match up to the result with the conventional CMOS-RPLA by using the simulator TINA-PRO. The proposed low power reversible PLA be able to implement numbers of Boolean functions like adder/subtractor and it can be transformed to perform desired function. The transient response or the spikes generation in the proposed design is also discussed.

## 2. BACKGROUND

This section provides the complete background of the technology used in the proposed design. This paper shows the physical design of the RPLA constituting by two major technologies: Reversible logics and Power gating.

### 2.1. Overview of Reversible logics

This part of the section 2 provides the brief of the reversible gates used in the reversible programmable logic design. According to the theory of the reversibility in the digital logics, the logic must have the equal number of inputs and outputs and they must be bijective. That means, an p×q of reversible gate consisting of p inputs and q outputs with design of each input assignment to a unique output assignment and vice versa OR there is one- to- one mapping from the inputs to the outputs and vice- versa[5] and p=q [4]. There are varieties of the reversible gates like Feynman gate (FG), Toffolli gate (TG), Fredkin gate (FRG), Peres gate (PG), New gate (NG, MKG, HNG and TSG, MG) [3]. Without discussing the all reversible gates, this paper describe only the MUX gate and Feynman gate as both gates with changed configuration are used in the RPLA.

### 2.1.1 Feynman Gate

The 2×2 reversible gate shown in Fig.2 called Feynman gate [2]. This gate is also documented as controlled- not gate (CNOT) having two inputs (A, B) and two outputs (P, Q). The outputs are defined by P=A, Q=A XOR B .To copy a signal, this gate can be used in the design to remove fan-out as fan-out is not permitted in reversible logic circuits. Quantum cost of a Feynman gate is 1.

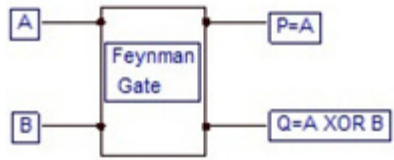

Fig.2. 2×2Feynman gate structure

### 2.1.2 MUX Gate

The pictorial representation of 3×3 reversible gate MUX (MG) gate is shown in Fig.3 [3]. This gate is a conservative gate having three inputs (A, B, C) and three outputs (P, Q, R). The outputs of the gate are defined by P=A, Q=A XOR B XOR C and R= A'C XOR AB. Its Quantum cost is 4.

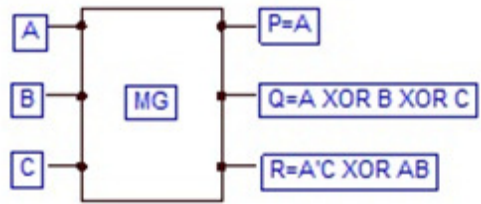

Fig.3: 3×3 MUX gate

So, these are the reversible logic gates which have been used in the RPLA design at the gate level implementation for reducing power consumption.

## 2.2 Overview of Power gating

This part of the section 2 provides the useful detail to understand the factors by which the power is dissipated in the systems and the power gating technique to reduce sub threshold leakage.

There are PMOS and NMOS transistors in the CMOS (Complementary metal oxide semiconductor) design of the circuits as pull-up and pull-down and both the transistors contributing equally in the circuit operation [5] & the major cause of power dissipation in the digital circuits are following [5][6][7][10]:

Dynamic or switching power consumption due to capacitance charging and discharging.
Short circuit power consumption due to energy require to charge up the parasitic load capacitance in the circuit.
The Leakage power consumption :
Static power consumption which is due to circuit present in the system which is other than the CMOS and providing a current flow from power supply to ground constantly.
The total power dissipation in CMOS digital circuits can be expressed as the sum of four components,

$$P_{avg.} = P_s + P_{s.c.} + P_{leakage} = \alpha_{0 \to 1} C_l . V^2_{dd} . f_{clk} + I_{s.c.} . V_{dd} + I_{leakage} . V_{dd} + I_{static} . V_{dd}$$

In the above equation $C_l$ is the load capacitance, $f_{clk}$ is the clock frequency, $\alpha_{0 \to 1}$ is the probability that a power consuming when transition occurs, $I_{s.c.}$ is the short circuit current (when both nmos and pmos active), $I_{leakage}$ is the leakage current arises from sub threshold effects.

### *Sleep Transistor*

Sleep transistor uses in the design is one of the emergent techniques to reduce the subthreshold leakage at the time of inactivity of the system OR when the system is not operation [12] [15]. Here the term, Sub threshold leakage which is used for one of the power dissipation whose cause is the carrier diffusion between the source and the drain region of the transistors in weak inversion and the amount of the sub threshold current might become considerable when the gate to source voltage is slighter than but very close to the threshold voltage of the device [12]. A sleep transistor is generally may be of PMOS or NMOS but with the high threshold voltage ($V_{th}$) value. So, the sleep transistor is a PMOS or NMOS transistor of high ($V_{th}$)which is placed in series with a low $V_{th}$ device unit. In the power gating design, circuit operates in two different modes [12]

### *Active Mode*

In this mode, the sleep transistor used in the design is turned ON and acts as the functional redundant resistance

*Sleep Mode*

This is the mode, in which the sleep transistor is turned OFF to reduce dynamic and leakage power in the standby mode.

When a sleep transistor is placed near Vdd, then called header switch and when a sleep transistor put near to the ground, called footer switch as shown in fig.5 (a), fig.5 (b).

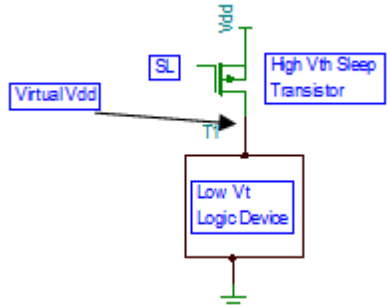

Fig 5(a) Header Switch

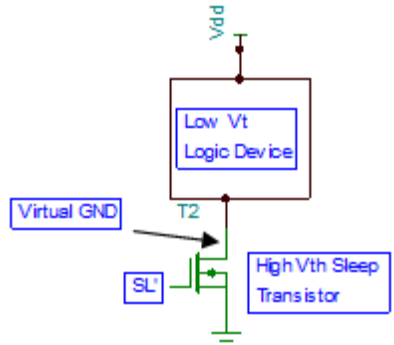

Fig.5 (b) Footer Switch

The figures are for the header and footer configurations and the PMOS is used for implementing the header switch putted close to Vdd for controlling the supply and the NMOS is to control GND, putted close to GND. The PMOS transistor is a smaller amount of leaky than the NMOS transistor and PMOS also has lower drive current than the NMOS transistor [15]. Due to these situations, PMOS takes more area than the NMOS which is not an interest or expectation for the very large scale integration. So, with these benefits of the NMOS i.e. footer switch of high drive current & smaller area over the PMOS, this paper considering the footer switch only for the proposed design.

## Analysis of Footer Switch

This particular subsection describes the connection between ground voltage and the leakage saving. For the relationship, a footer switch same as shown in fig 5(b) is taken which is biased in the weak inversion i.e. $V_g < V_{th}$ [16]. the leakage of the single transistor provides the leakage of the logic circuit [11].

So,

$$I_{leakage} \text{ (circuit)} = I_{leakage}(\text{Footer})$$

But

$$I_{leakage} = I_o \left(\frac{W}{L}\right) 10^{\frac{(v_{g-}v_{th}) + \eta(v_{ds})}{ss}} \text{ for the logic circuit}$$

...............(1)

So, equation becomes

$$I_o \left(\frac{W \text{ cicuit}}{L}\right) 10^{\frac{(-Vthc) + \eta(V_{dd}-Vgnd)}{ss}} = I_o \left(\frac{WFooter}{L}\right) 10^{\frac{(v_{g-}v_{thF}) + \eta(v_{ds})}{ss}}$$

.............. (2)

Vthc and VthF shows the threshold voltage of the logic circuit and the footer device resp., $\eta$ is the Drain induced barrier lowering ( a secondary effect) coefficient and ss is the sub threshold slope.

After solving eq 2.

$$V_{gnd=} \frac{-V_g + s_s \log_{10}\left(\frac{Wcircuit}{Wfooter}\right) + (Vthf - Vthc + \eta Vdd}{2\eta}$$ ........(3)

The equation 3 shows that Vgnd is proportional to the gate voltage Vg with negative slope. So, if the footer gate voltage is increased, there is decrease in the ground potential and vice-versa.
To make the trade-off between leakages saving and wakeup- overhead, the control over Vgnd should be needed. So,

$$\frac{I \text{ Sleep}}{I \text{ active}} = 10^{-\left(\eta \frac{(Vdd-Vgnd)}{ss}\right)}$$ .....(4)

So, by the above terms, higher Vgnd results in higher leakage saving.

## 3. PROPOSED IMPLEMENTATION

The RPLA consists of Reversible AND array and Reversible OR arrays shown in fig. 6

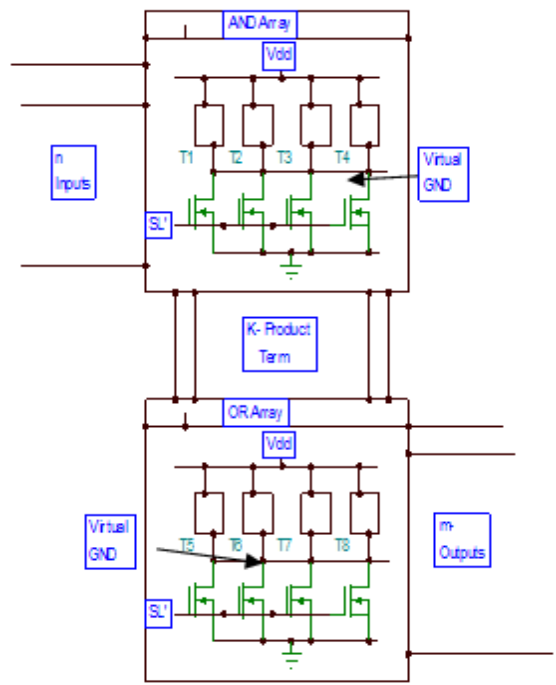

Fig 6 Proposed RPLA with Sleep Transistors

This architecture shows the Reversible AND array having n numbers of inputs which will provide the k product term. The product terms are fed to the Reversible OR array as its inputs and provide the sum of product. Sleep transistors are used in the internal structure of each array with footer configuration and by this architecture the user can easily vary this structure according to the expected Boolean function at any time. In this work, the example of 3- input PLA is taking into consideration for which three inputs are provided to the Reversible AND array and this will provide us eight- min terms or product terms. The Reversible OR gate will perform for the 8- min terms which again provide the sum of those.

### 3.1 Design of RPLA by sleep transistors

For the proposed design the TINA-PRO simulator is used. TINA Design suit is a great yet reasonable software package for analyzing, designing and real time testing of analog, digital, VHDL, and mixed electronic circuits and their layouts. For the simplicity, the macro of the each component of the RPLA is designed by using the power gating structure. The RPLA has two basic plane i.e. Reversible AND Plane and OR Plane. These two planes consisting of AND OR structures of the reversible gates respectively. The reversible AND plane of the RPLA consists of Feynman and MUX gates and reversible OR plane consists of MUX gate only. These gates contain the NOT, AND XOR functions. So, in starting the Macros of each gate is defined and then connected in a predefined manner. The complete Macros of the Feynman gate and MUX gate is shown in fig. 7.1 & 7.2 respectively.

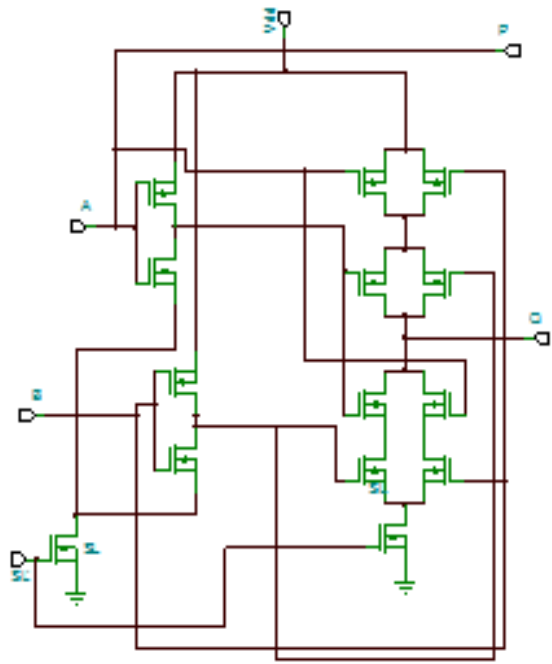

Fig. 7.1 Macro of Feynman gate with sleep transistor

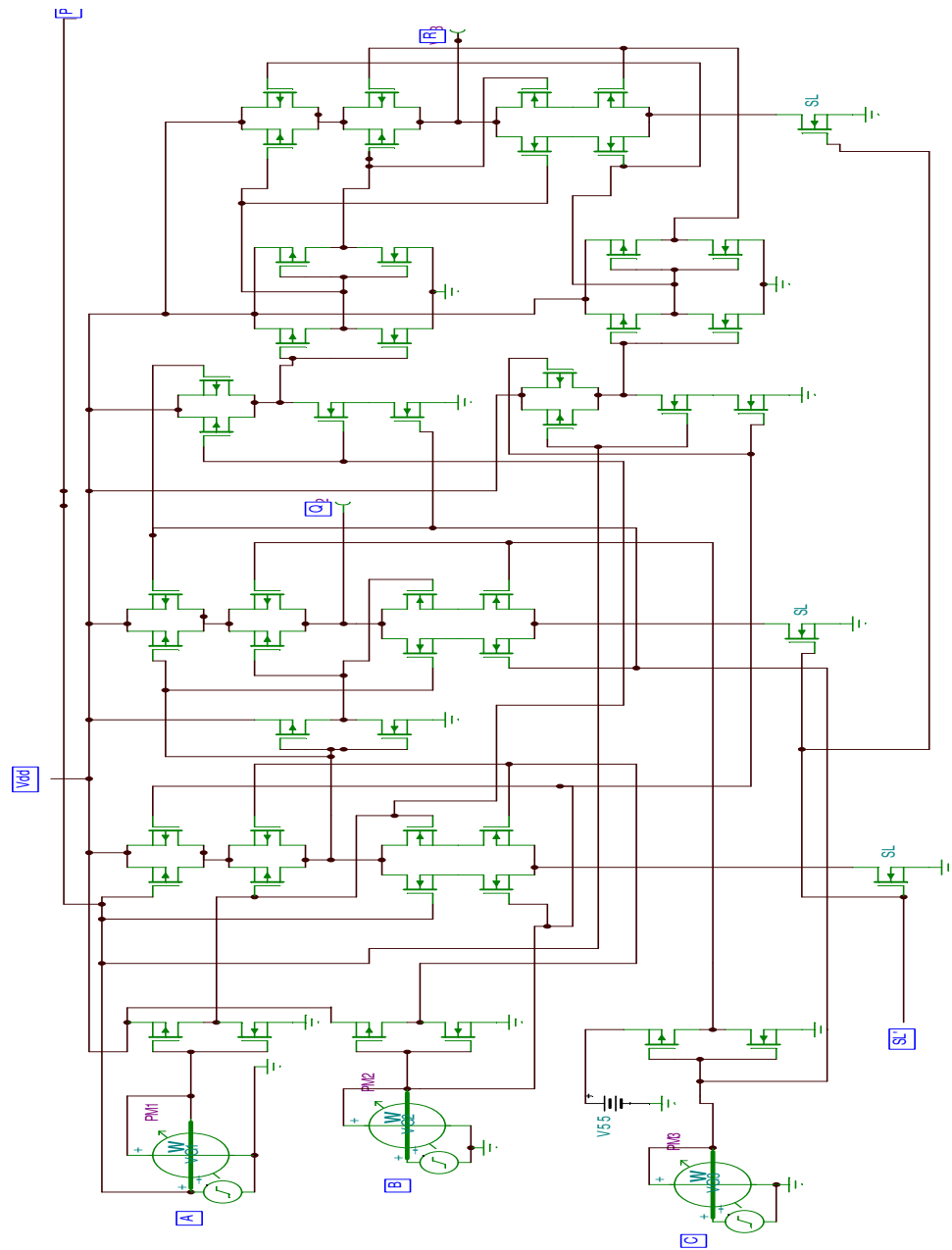

Fig. 7.2 Macro of MUX gate with sleep transistors

## 4. POWER CONSUMPTION MEASUREMENT

The wattmeter is used for measurement of the power dissipation in the array. VF1-VF8 is the output voltages and PM1- PM3 is power measurement wattmeter. The fig.8 (a) shows the complete structure of power measurement of Reversible AND array. And fig.8 (b) for reversible OR array.

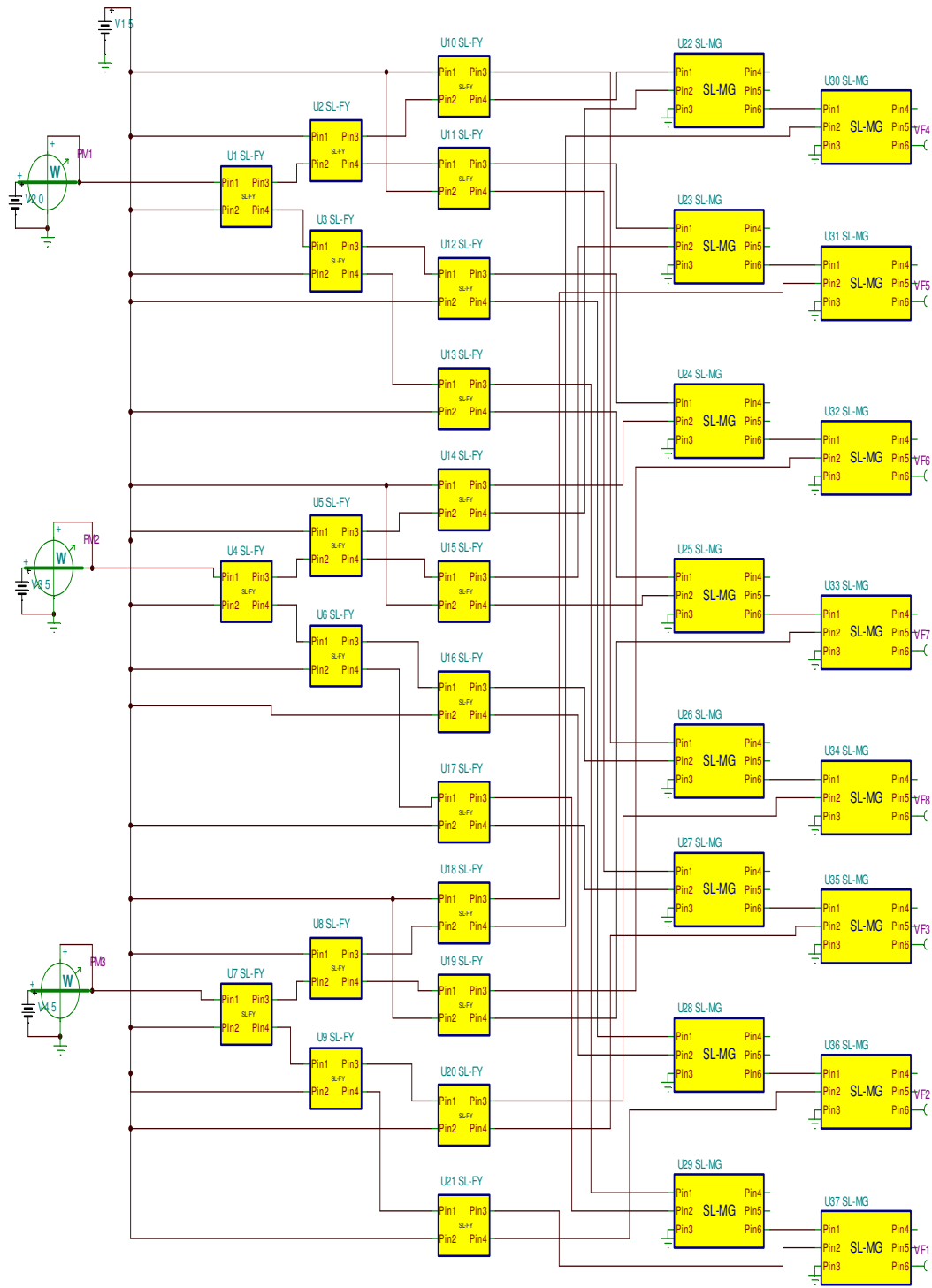

Fig. 8(a) Power measurement setup for Reversible AND array

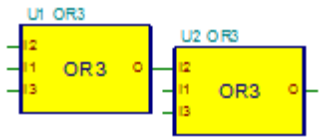

Fig. 8(b) Power measurement setup for Reversible OR array

## 5. RESULTS

This section describes the analysis of the performance of the proposed design using TINA- PRO. For analyzing the parameters like power consumption and transient response, Let consider the input values A, B, & C are at logic 1 or +5V unit step.

### 5.1 Power Analysis

For the power analysis, comparison of the results for the sleep transistors RPLA with the RPLA designed by the conventional CMOS technology is shown in table 1 below.

Table 1 Power consumption comparison between CMOS-RPLA & SL RPLA

| Input Vector ABC | Power Consumed(pW) | | | | | |
|---|---|---|---|---|---|---|
| | CMOS- RPLA | | | Proposed SL- RPLA | | |
| | PM1 | PM2 | PM3 | PM1 | PM2 | PM3 |
| 000 | 0 | 0 | 0 | 0 | 0 | 0 |
| 001 | 0 | 0 | 221.91 | 0 | 0 | 90.57 |
| 010 | 0 | 221.92 | 0 | 0 | 90.57 | 0 |
| 011 | 0 | 221.92 | 221.91 | 0 | 90.57 | 90.57 |
| 100 | 187.71 | 0 | 0 | 90.57 | 0 | 0 |
| 101 | 187.71 | 0 | 221.91 | 90.57 | 0 | 90.57 |
| 110 | 187.71 | 221.92 | 0 | 90.57 | 90.57 | 0 |
| 111 | 187.71 | 221.92 | 221.91 | 90.57 | 90.57 | 90.57 |

The PM1, PM2 & PM3 are the three watt meters connected on the input side at three inputs line (ABC). From the table we can see that when the input vector is at logic level 1 or high then the power consumption In the CMOS- RPLA is 187.71- 221.92 and the power consumption in the RPLA designed by the proposed technology is 90.57. From the analysis table it is apparent that there is a 40.8% saving of energy by using the power gating in the design.

## 5.2 Transient Response Analysis

"Transients", a term we'll use for simplicity here, are actually "Transient Voltages". More familiar terms may be "surges" or "spikes". Basically, transients are momentary changes in voltage or current that occurs over a short period of time. Generally in the systems, there are two kinds of analysis:

*The DC analysis*: Which tells Vout if Vin is constant?
*Transient analysis*: Which tells Vout(t) if Vin(t) changes.
This analysis used to measure the behavior of the circuits at the time of switching.
The comparable Transient response of the Conventional CMOS RPLA and Proposed SL- RPLA are shown below.

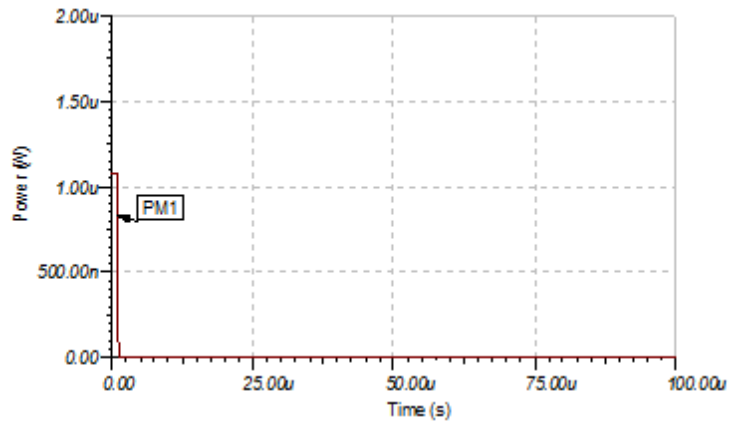

Fig. 9 (a) Transient Response of the Line A for CMOS- RPLA

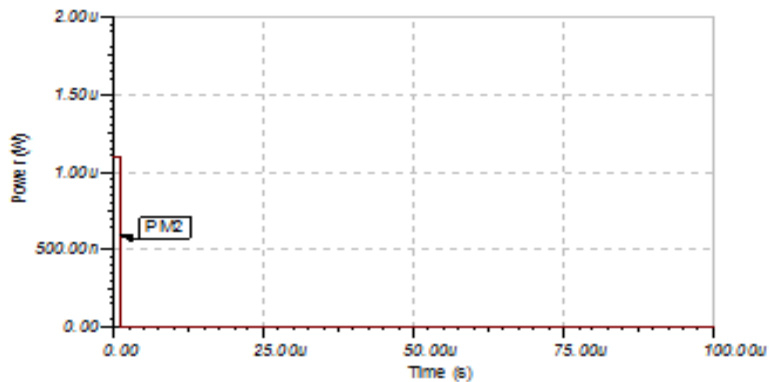

Fig. 9 (b) Transient Response of the Line B for CMOS- RPLA

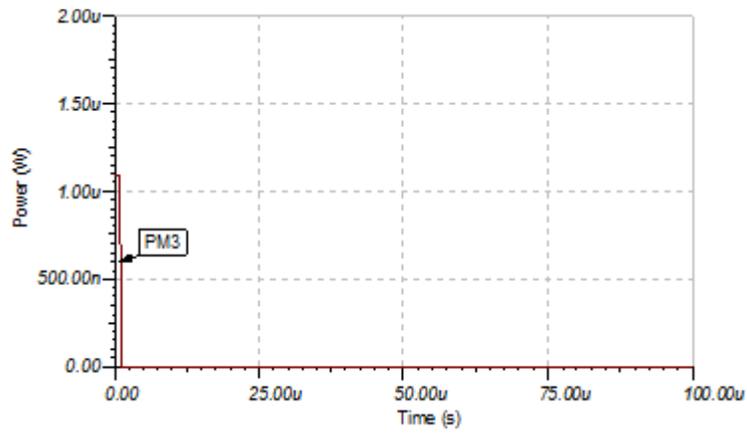

Fig. 9(c) Transient Response of the Line C for CMOS- RPLA

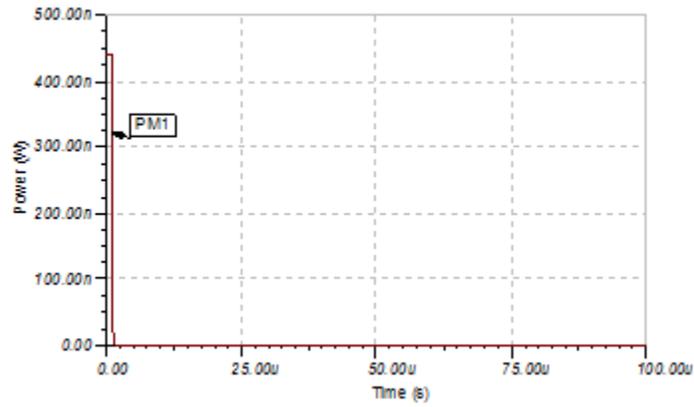

Fig. 9(d) Transient Response of the Line A for Proposed- RPLA

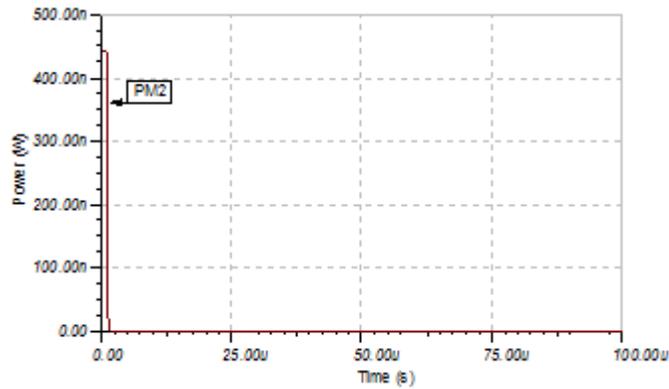

Fig. 9(e) Transient Response of the Line B for Proposed- RPLA

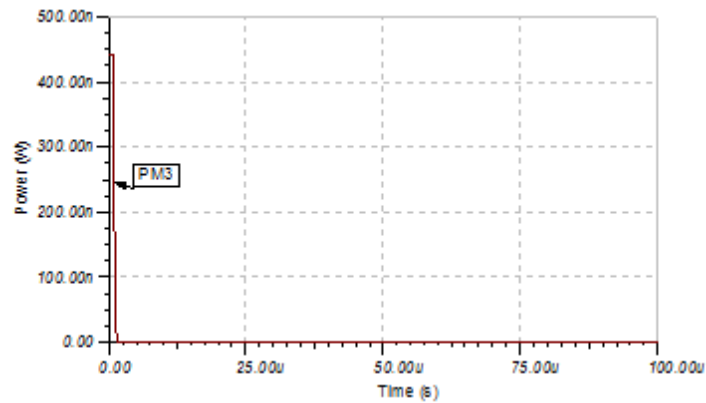

Fig. 9(f) Transient Response of the Line C for Proposed- RPLA

From the Power consumption measurement results of proposed technology against the conventional CMOS technology, it is clear that the sleep transistor saves the approx. 40.8% of the energy and from the transient response of the different lines of the proposed RPLA against CMOS- RPLA, it is conclude that the spikes generates at the time of very short period is approx. 440nW for the SL-RPLA structure and for CMOS- RPLA it is 1.1 uW. So, the less amplitude spike generates means more stability in the system.

So, by these transient responses and the power consumption measurement, it is clear that the proposed technology of the multi-threshold transistor or Power gating for the reversible programmable logic array design is superior and efficient for saving the energy which is dissipated by the sub-threshold leakage at the time of inactivity.

## 6. CONCLUSION

Reversible computing is one of the powerful technologies to reduce power consumption at the gate level implementation but at the physical level design, in nanometer scale CMOS technology, sub threshold leakage power is also one of the great challenge in front of the circuit designer. This paper, presented a new circuit design named "Power gating structure" to tackle the leakage problem which uses the sleep transistors.  In this paper, the complete concept of the power gating with the footer switch has been explained and used this concept for the designing of RPLA at the transistor level. The design architecture of the RPLA at the transistor level is done by TINA- 8 software. The power consumption comparison for both the technique i.e. CMOS- RPLA and the SL- RPLA has been also shown in the paper and showing the results for the proposed technique is better than the CMOS technique. The proposed technique for the physical level design of RPLA saves the 40.8% of the power consumption as compares to the CMOS technology. For the analyzing if the switching activity, the transient response has also been shown for the proposed RPLA. The transient response results for the proposed RPLA show the spikes produced by the proposed design are less than the CMOS – RPLA. The spikes generates for the proposed MTCMOS- RPLA is of 440nW which is less comparable to the CMOS-RPLA having 1.20uW.